\documentclass[aps,prc,twocolumn,superscriptaddress,showkeys,preprintnumbers]{revtex4-2}

\usepackage[utf8]{inputenc}
\usepackage[T1]{fontenc}
\usepackage{amsmath,amssymb,bm}
\usepackage{graphicx}
\usepackage{booktabs}
\usepackage{siunitx}
\usepackage{xcolor}
\usepackage{microtype}
\usepackage{placeins}

\usepackage{hyperref}
\hypersetup{colorlinks=true,linkcolor=blue,citecolor=blue,urlcolor=blue}

\usepackage[normalem]{ulem}

\begin{document}

\title{CLVisc Agent for autonomous relativistic hydrodynamics studies}

\author{Qi Wang}
\affiliation{Key Laboratory of Quark and Lepton Physics (MOE) \& Institute of Particle Physics, Central China Normal University, Wuhan 430079, China}
\affiliation{Artificial Intelligence and Computational Physics Research Center, Central China Normal University, Wuhan 430079, China}

\author{Long-Gang Pang}
\email{lgpang@ccnu.edu.cn}
\affiliation{Key Laboratory of Quark and Lepton Physics (MOE) \& Institute of Particle Physics, Central China Normal University, Wuhan 430079, China}
\affiliation{Artificial Intelligence and Computational Physics Research Center, Central China Normal University, Wuhan 430079, China}

\author{Shi Pu}
\email{shipu@ustc.edu.cn}
\affiliation{Department of Modern Physics and Anhui Center for fundamental Sciences
(Theoretical Physics), University of Science and Technology of China,
Anhui 230026}
\affiliation{Southern Center for Nuclear-Science Theory (SCNT), Institute of Modern
Physics, Chinese Academy of Sciences, Huizhou 516000, Guangdong Province,
China}

\author{Xin-Nian Wang}
\email[]{xnwang@lbl.gov}
\affiliation{Key Laboratory of Quark and Lepton Physics (MOE) \& Institute of Particle Physics, Central China Normal University, Wuhan 430079, China}
\affiliation{Artificial Intelligence and Computational Physics Research Center, Central China Normal University, Wuhan 430079, China}

\begin{abstract}
We enable large language model (LLM) agents to autonomously perform end-to-end hydrodynamic simulations of the quark–gluon plasma evolution and calculation of final hadron spectra in relativistic heavy-ion collisions. We design a meta skill that allows an agent to explore a project’s source code, craft a specialized skill, and iteratively refine it. Applying this meta skill to the (3+1)D viscous hydrodynamic code CLVisc, the agent builds a CLVisc skill encoding its operational knowledge and then independently executes full scientific workflows: designing parameter scans, running simulations, comparing ensemble results, and producing publication-ready figures. Crucially, the agent draws on literature-informed heavy-ion physics to select physically meaningful observables and interpret outcomes without explicit instruction. We demonstrate the pipeline in two scenarios: temperature-dependent shear viscosity over entropy density $\eta/s$, and nuclear-structure effects in O+O collisions at $\sqrt{s_{\mathrm{NN}}} = 5.36$~TeV using four \textit{ab initio} descriptions of $^{16}$O. In both, the agent plans, executes, and analyzes autonomously, devising new initial-state observables to explain final observations and extract qualitative knowledge. The meta skill is agnostic to code versions and Monte Carlo generators, promising future multi-agent systems in high-energy nuclear physics.
\end{abstract}


\maketitle

\section{Introduction}
\label{sec:intro}

The application of artificial intelligence in high-energy nuclear physics \cite{He2023HENPML,Boehnlein2022RMP} has traditionally centered on four well-defined pillars: discrimination (e.g., classifying QCD phase transition) \cite{Pang2018EOSMeter,Pang2019QCDTransition}, generation (e.g., flow model or diffusion model for fast event generators)~\cite{Gao2020NormalizingFlows,Mikuni2023Diffusion}, representation (e.g., physics-informed neural networks)~\cite{Raissi2019PINN,Dai:2026tjl,Dai:2026nzp}, and operation (e.g., detector control)~\cite{Edelen2024AcceleratorControl,Kvapil2025RealtimeAI}. Yet, the advent of large language models (LLMs) and, more critically, autonomous LLM agents---systems capable of iterative planning and tool use~\cite{Yao2023ReAct,Schick2023Toolformer,Ren2025ScientificAgents,bran2023chemcrow}---has remained conspicuously absent from this landscape. While LLMs have begun to permeate other scientific domains as "AI scientists" \cite{Lu2024AIScientist,Yamada2025AIScientistV2,Zheng2025LLMScientificDiscovery, Boiko2023Coscientist,AlphaEvolve2025,schmidgall2025agent}, and recent benchmarks have started
to evaluate their ability to perform scientific reasoning, research coding, and end-to-end paper reproduction \cite{ScienceAgentBench,SciCode,PhyBench,PRBench,qiu2026end,NDUM2025100555,ni2024mechagents,li2026autonomous,Moreno:2026mqk,Amram:2024fjg,Zhang:2024kws,Fanelli:2024ktq,Heneka:2025fpe,McGreivy:2025rrz,Rafique:2025xeo,Mermer:2025mcf,Song:2025odk,Bakshi:2025fgx,Diefenbacher:2025zzn,Jat:2026yxv,Mallampalli:2026hrl,Tan:2026ier,Plehn:2026gxv,Knipfer:2026kng,Badea:2026klb,Agrawal:2026lvg,Menzo:2026qrl,Moreno:2026mqk}, their potential to navigate the complex, multi-stage workflows of heavy-ion collision phenomenology remains entirely unexplored. This gap represents a missed opportunity to automate the labor-intensive parameter scans that are essential for constraining the transport properties and initial geometry of the quark-gluon plasma (QGP).

Extracting QGP properties from heavy-ion collisions---
such as the equation of
state and the QCD phase transition, the vorticity and spin polarization of the
medium~\cite{Liang:2004ph, Liang:2004xn, STAR:2017ckg, Becattini:2024uha}, and its transport coefficients, most notably the
specific shear and bulk viscosities $\eta/s$ and $\zeta/s(T)$---requires systematic comparisons between event-by-event (3+1)D viscous hydrodynamic simulations and experimental data \cite{Bernhard2019Bayesian,JETSCAPE2021MultiSystem,Parkkila2021Bayesian,Jia2024NuclearShapeImaging,PhysRevLett.106.212302,PhysRevLett.116.212301,Zhang_2019}.
Here, we focus on two representative aspects:  the temperature-dependent shear viscosity over entropy density  $\eta/s$ and the imprints of nuclear structure. Traditional workflows, however, remain stubbornly manual: a researcher must hand-edit input configurations, submit batch jobs, parse large volumes of event-by-event output, and aggregate final-state observables \cite{Shen2016iEBEVISHNU,Karpenko2014vHLLE,Schenke2010MUSIC}. This procedure is not only labor-intensive but fundamentally restricts the dimensionality of the parameter space that can be feasibly explored \cite{Novak2014Bayesian,Bernhard2016Quantifying,Bernhard2019Bayesian}.

This traditional manual workflow, despite its inefficiency, offers a distinct advantage over modern Bayesian analysis: researchers accumulate qualitative intuition regarding how specific physical parameters influence final-state observables through iterative, hands-on tuning to match experimental data. In contrast, Bayesian global analyses aggregate vast datasets to extract correlated posterior distributions of model parameters \cite{Novak2014Bayesian,Bernhard2016Quantifying,Bernhard2019Bayesian,JETSCAPE2021MultiSystem,Nijs2021Trajectum}, but they often obscure the direct causal relationship between a single physical quantity and a particular observable---the underlying patterns remain buried within the high-dimensional likelihood landscape. Standard metrics, such as the Pearson correlation coefficient, capture only linear dependencies and are thus ill-suited to characterize the complex, non-equilibrium dynamics of heavy-ion collisions.

Here, LLM agents offer a novel opportunity to bridge this gap. Manually extracting such targeted qualitative knowledge by sifting through thousands of event-by-event hydrodynamic simulations with varying parameter configurations would be prohibitively laborious. However, an LLM agent is uniquely suited to this task: it can execute the planning, acting, and analyzing loop required to compare ensemble outcomes and distill, in a non-linear manner, which parameters govern which observables \cite{Yao2023ReAct,Schick2023Toolformer,Ren2025ScientificAgents,Zhang2025LLMScience,SciCode,PRBench}. In this way, the agent transforms what would otherwise be an insurmountable manual effort into a scalable, automated discovery process.

In this work, we present a end-to-end framework for agentic knowledge discovery that automates the full research pipeline for relativistic hydrodynamics studies. We developed a SKILL---a structured knowledge module---that encodes expertise for running CLVisc \cite{Pang:2018zzo,Wu:2021fjf}, a GPU-accelerated (3+1)D viscous hydrodynamics code. Using the kimi-cli agent platform, we demonstrate how this SKILL enables an LLM to independently conduct systematic parameter scans, analyze results, and generate quantitative summaries of parameter--observable relationships.

To validate the generality and scientific utility of our approach, we apply this agentic framework to two distinct physics questions. \textbf{Scenario~I} investigates the effects of temperature-dependent shear viscosity  over entropy density on heavy-ion observables for Pb$+$Pb collisions at $\sqrt{s_{\mathrm{NN}}} = 5.02$~TeV in the 30--40\% centrality. \textbf{Scenario~II} applies the same automated pipeline to O$+$O collisions at $\sqrt{s_{\mathrm{NN}}} = 5.36$~TeV, comparing the hydrodynamic response obtained from four distinct \textit{ab initio} nuclear structure models. Our goal is twofold: first, to demonstrate that a data-rich, automated scanning framework can reveal subtle but statistically robust differences that would be difficult to access through ad hoc manual comparisons; and second, to establish a reproducible platform for future AI-assisted or LLM-guided searches for novel observables.

This paper is organized as follows. Section~II describes the agentic workflow architecture, the CLVisc $(3{+}1)$D viscous hydrodynamics model, and the physical setup for the two scenarios. Section~III presents the results of both scans: the temperature-dependent $\eta/s$ scan in Pb+Pb collisions and the nuclear-structure comparison in O+O collisions. We conclude with a discussion and conclusion in Sec.~IV.

\section{Methods}
\label{sec:methods}

\subsection{Agentic workflow architecture}
\label{sec:workflow}


We developed an automated research pipeline powered by LLM agents, designed to perform simulations of high-energy heavy-ion collisions using relativistic hydrodynamics (CLVisc). In contrast to traditional workflows that couple multiple Monte Carlo models written in different languages (C++/Fortran/Python) and require experts to jointly develop a uniform interface for data transfer—as exemplified by the JetScape collaboration \cite{putschke2019jetscape,JETSCAPE:2020shq,JETSCAPE:2020mzn} —the agentic workflow possesses detailed knowledge of each Monte Carlo model, including how to modify its configuration files and how to pipe data from one code to another. This makes hybrid simulations quick and easy for graduate students new to the field, thereby accelerating discovery.

The core of this agentic workflow is the LLM agent. Unlike a standard LLM, which merely communicates with users via natural language messages, an agent is further equipped with designed prompts that assign it a role, specific abilities, and the capacity to reflect, to reason using chain-of-thought, to access (read/write) files in a local directory, to write and execute Bash or Python scripts, to inspect runtime error messages, and to iteratively debug and re-execute the code until correct. Remotely, many tools are made accessible to agents through the Model Context Protocol (MCP). The LLM agent has become a pivotal concept in artificial intelligence, poised to boost productivity across many domains—including high-energy nuclear physics.

\begin{figure*}[htbp]
    \centering
    \includegraphics[width=\linewidth]{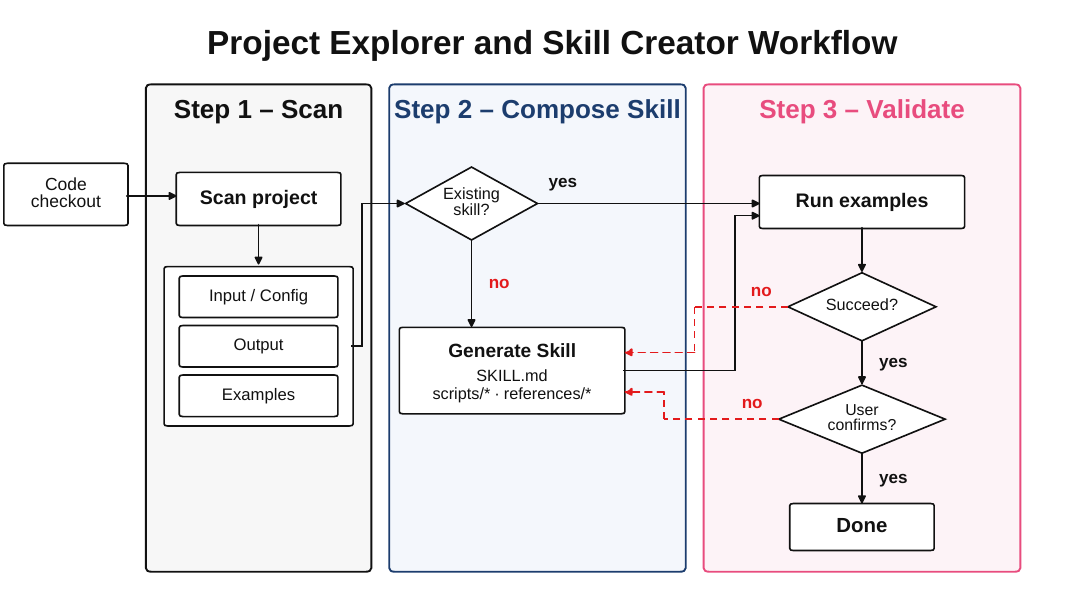}
    \caption{The scan--compose--validate workflow of the
\texttt{project-explorer-and-skill-creator} meta-skill. A model-agnostic engine
first scans a given scientific-code checkout and builds a project profile
capturing its inputs and configuration, outputs, and runnable examples
(Step~1). If a skill already exists for the checkout, the workflow proceeds
directly to validation; otherwise a version-specific skill is generated,
comprising \texttt{SKILL.md}, executable scripts, and references, with launch
commands derived from the README and source code (Step~2). The skill is then
validated by running its examples: if the run fails or the user does not
confirm the result, the workflow loops back to regenerate the skill; once the
run succeeds and is confirmed, the workflow terminates (Step~3).}
\label{fig:skillcreator}
\end{figure*}

To maximize the capabilities of LLM agents, we now commonly provide them with additional “ SKILLs.” A SKILL is analogous to the domain knowledge of a human expert specialized in solving a particular type of problem, and also resembles the concept of a standard operating procedure (SOP) in large organizations. A SKILL is usually delivered as a self-contained folder with several files or subfolders. The central file is SKILL.md, whose first few lines briefly describe the skill’s name, purpose, and the contexts in which the agent should apply it. When a user makes a request in natural language, the agent first searches for relevant skills using these descriptions. The remainder of the SKILL.md file contains detailed information on workflows, successful examples, common pitfalls, scripts the agent can use, and reference documents. This structure automatically explains the purpose of the other subfolders: typically, one subfolder holds scripts that the agent can execute to produce reliable outputs, and another contains useful documents such as references or formatting standards. Claude Code \cite{ClaudeCode2025}, for example, offers several scientific research skills that one can explore. In the scientific-visualization-skill, for instance, the example scripts are Python programs for generating publication-quality figures, and the reference files document the distinct formatting requirements of various journals (e.g., Nature, Science, and APS journals). The SKILL ensures reproducibility and enables the agent to execute complex workflows without extensive retraining or fine-tuning.

We implement this workflow in \texttt{kimi-cli} \cite{KimiCLI2025}, which provides the agent with
access to local files, scripts, and command-line execution. While a CLVisc-specific skill can be created interactively from a local installation,
applying this procedure reliably across many versions and forks of a
simulation code---which may differ in entry points, argument conventions,
configuration formats, and output layouts---motivates a more systematic
approach. We therefore developed a dedicated meta skill, \texttt{project-
explorer-and-skill-creator}, that automates skill creation as a reproducible
\emph{scan--compose--validate} workflow  also see Fig.~\ref{fig:skillcreator}.

The meta skill is organized as a generic engine plus detachable knowledge
packs. The engine is model-agnostic: it scans a local checkout and extracts
only structural facts, including entry points, parameters, output files, build
targets, and any run commands documented in the project README. Model-family
knowledge is supplied separately through thin per-model knowledge packs, which
provide the physical meaning of parameters and outputs, observable extractors,
and the minimal run-profile information needed to launch a job. In this way,
model-specific operational knowledge remains explicit and modular, while the
engine itself remains independent of any particular simulation code. Extending
the framework to another code family, such as SMASH or URQMD, therefore
requires only the addition of a new knowledge pack rather than modification of
the engine.

The workflow is organized into three main modules: \emph{Scan}, \emph{Compose
Skill}, and \emph{Validate}. In the \emph{Scan} module, the engine first scans
the source tree and then builds a project profile, summarizing entry points,
parameters, configuration files, output files, build targets, README-documented
commands, version fingerprints, and unresolved gaps. In the \emph{Compose Skill}
module, the engine first checks whether a usable skill already exists for the
checkout; if so, the workflow proceeds directly to validation, and otherwise it
assembles a version-specific skill with the same
\texttt{SKILL.md}/\texttt{references}/\texttt{scripts} layout described above.
The launch command is derived rather than hardcoded: documented commands are
first extracted from the README, then cross-checked against the argument
conventions required by the source, and finally reconstructed from the code
itself when the documentation is incomplete or stale. In the \emph{Validate}
module, the generated skill is exercised through tiered real execution. The
validation first checks the runtime environment and required components, then
attempts a real run to the first non-empty output, and finally extracts one
representative observable and produces a diagnostic plot. The deepest tier
actually reached is recorded explicitly in a verification report, and blocked
tiers are never reported as success. Missing dependencies, failed builds, or
unavailable external modules are reported as blockers unless the user explicitly
authorizes repair. If the run fails, or if the user does not confirm that the
resolved run flow and the extracted observable are correct, the workflow loops
back: the relevant knowledge pack is refined and the SKILL.md is regenerated and re-validated. This iteration continues until a verified run is obtained and
confirmed, before production runs are launched.

In the present work, the CLVisc knowledge pack encapsulates several layers of
domain expertise. At the operational level, it specifies the simulation chain,
including hydrodynamic evolution, hadron production, and the extraction of
final-state observables. At the conceptual level, it provides the physical
interpretation of key parameters such as the initial proper time $\tau_0$, the
specific shear viscosity $\eta/s$, the bulk viscosity $\zeta/s$, and the
freeze-out temperature $T_{\mathrm{frz}}$. At the analytical level, it
supplies observable extractors and default expectations for how variations of
transport parameters should influence selected observables. Comparing
extracted results against these expectations enables the agent to test
hypotheses and identify potentially informative deviations.

When given a specific research objective, the agent first designs a parameter-scan strategy, then consults the generated skill to determine the required
execution steps, produces scripts to automate the workflow, and systematically
carries out the simulation and analysis. After completing the Monte Carlo
runs, the agent analyzes event-by-event results using conventional observables
and may also propose additional observables to test its working hypotheses.
This architecture decouples high-level scientific objectives from low-level
implementation details, enabling researchers to formulate physics questions in
natural language while the agent handles the complexity of execution,
validation, and analysis.

\subsection{Viscous hydrodynamics with CLVisc}
\label{sec:clvisc}

The space-time evolution of the QGP medium is simulated with CLVisc~\cite{Pang:2018zzo,Wu:2021fjf}, 
a GPU-accelerated (3+1)D relativistic viscous hydrodynamic framework 
designed for event-by-event heavy-ion collision simulations. Starting from 
fluctuating initial energy-density profiles, CLVisc evolves the energy-momentum 
tensor according to local conservation laws,
\begin{equation}
    \nabla_\mu T^{\mu\nu}=0,
\end{equation}
where the energy-momentum tensor is decomposed as
\begin{equation}
    T^{\mu\nu}
    =
    e u^\mu u^\nu
    - (P+\Pi)\Delta^{\mu\nu}
    + \pi^{\mu\nu}.
\end{equation}
Here, $e$ is the local energy density, $P$ is the equilibrium pressure, 
$u^\mu$ is the fluid four-velocity, $\Pi$ denotes the bulk viscous pressure, 
$\pi^{\mu\nu}$ is the shear-stress tensor, and 
$\Delta^{\mu\nu}=g^{\mu\nu}-u^\mu u^\nu$ projects onto the space orthogonal 
to the local flow velocity. The pressure is determined from the equation of 
state, which closes the hydrodynamic equations.

Viscous corrections are evolved within a causal Israel--Stewart-type theory. 
The shear and bulk viscous tensors relax toward their Navier--Stokes limits 
on finite relaxation time scales, thereby avoiding the acausal behavior of 
first-order viscous hydrodynamics. In the present work, the shear viscosity 
to entropy density ratio $\eta/s$ and the bulk viscosity to entropy density 
ratio $\zeta/s$ are treated as external transport inputs. Their values, or 
temperature-dependent parametrizations, are specified in the hydrodynamic 
configuration files and can be modified automatically by the LLM agent during 
parameter scans.

For each event, the initial entropy or energy density profile generated by 
the initial-state model is converted into a three-dimensional hydrodynamic 
profile at the starting proper time $\tau_0$. The subsequent evolution is 
performed until the medium cools down to a prescribed freeze-out temperature 
$T_{\mathrm{frz}}$. The freeze-out hypersurface is then converted into hadron 
spectra using the Cooper--Frye formula,
\begin{equation}
    E\frac{dN_i}{d^3p}
    =
    \frac{g_i}{(2\pi)^3}
    \int_\Sigma
    f_i(x,p)\, p^\mu d\Sigma_\mu ,
\end{equation}
where $g_i$ is the degeneracy factor of hadron species $i$, 
$d^3\Sigma_\mu$ is the hypersurface element, and $f_i(x,p)$ denotes the 
phase-space distribution including viscous corrections. The resulting final-state 
particles are used to compute identified-hadron spectra, mean transverse 
momenta, anisotropic flow coefficients, and correlation observables.

This hydrodynamic framework provides the common dynamical backbone for both 
physics scenarios considered in this work. In Scenario~I, the agent modifies 
the temperature dependence of $\eta/s$ while keeping bulk viscosity switched off 
in order to isolate the effect of shear viscous damping. In Scenario~II, the 
same hydrodynamic evolution and analysis chain are applied to different 
\textit{ab initio} nuclear-structure inputs, ensuring that variations in final-state 
observables can be attributed to the initial nuclear geometry rather than to 
changes in the medium evolution.

\subsection{Scenario I: Temperature-dependent shear viscosity over entropy density}
\label{sec:methods_shear}

To investigate the temperature dependence of shear viscosity, we instructed the agentic pipeline to perform a systematic parameter scan in Pb$+$Pb collisions at $\sqrt{s_{\mathrm{NN}}} = 5.02$~TeV in the 30--40\% centrality class. Initial conditions were generated using the T\raisebox{-.5ex}{R}ENTo model \cite{Moreland:2014oya} with a nucleon width of $w = 0.5$~fm and a gamma shape parameter $k = 1.0$. The hydrodynamic evolution starts at the initial proper time $\tau_0 = 0.6$ fm, and the system evolves until the local temperature drops to the freeze-out temperature $T_{\rm frz} = 0.137$ GeV. The HotQCD equation of state is adopted throughout the     evolution~\cite{Bazavov:2014pvz}. Once tasked, the agent autonomously modified the CLVisc configuration files, executed the GPU-accelerated hydrodynamic evolution, and extracted final-state observables for each parametrization without manual intervention at the per-event level.

\begin{figure}[htbp]
\centering
\includegraphics[width=\linewidth]{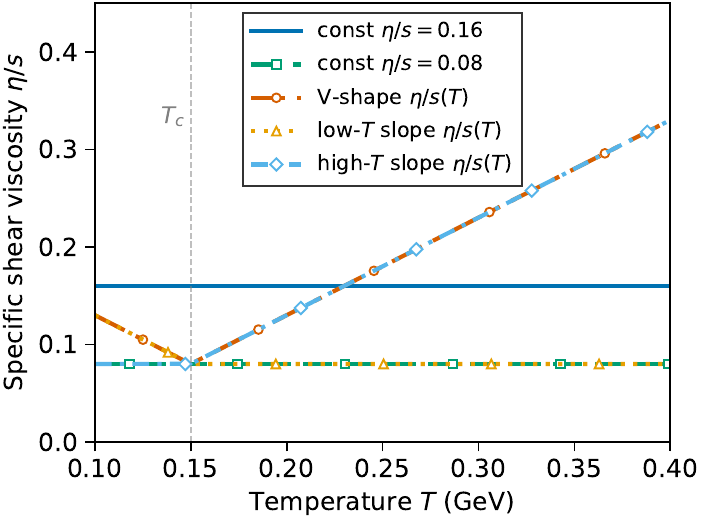}
\caption{Temperature-dependent shear viscosity $\eta/s$ parametrizations used in Scenario~I. The five curves represent: constant $\eta/s = 0.16$ (blue solid), constant $\eta/s = 0.08$ (green dashed with square markers), V-shaped with minimum at $T_c = 0.15$~GeV (orange-red dash-dot with circle markers), low-temperature slope only (orange dotted with triangle markers), and high-temperature slope only (light-blue dash-dot with diamond markers). The low-temperature-slope profile coincides with the constant $\eta/s=0.08$ curve for $T>T_c$, while the high-temperature-slope profile coincides with the V-shaped profile for $T>T_c$.}
\label{fig:eta_s_profiles}
\end{figure}

To isolate the effects of shear viscosity, bulk viscosity was completely turned off for all simulations. The temperature dependence of $\eta/s$ was implemented using CLVisc's piecewise-linear parametrization:
\begin{equation}
\frac{\eta}{s}(T) = \begin{cases}
\text{left\_slope} \times (T - T_\text{min}) + \eta_\text{min} & T < T_\text{min} \\
\text{right\_slope} \times (T - T_\text{min}) + \eta_\text{min} & T \geq T_\text{min}
\end{cases}
\end{equation}
where $T_\text{min}$ locates the minimum of the $\eta/s$ profile and $\eta_\text{min}$ specifies the viscosity at that point. The following five parametrizations were adopted (see Fig.~\ref{fig:eta_s_profiles}):
\begin{enumerate}
    \item[\textbf{(1)} ] \textbf{Constant high:} $\eta/s = 0.16$ (baseline with strong viscous damping)
    \item[\textbf{(2)} ] \textbf{Constant low:} $\eta/s = 0.08$ (baseline with weak viscous damping)
    \item[ \textbf{(3)} ] \textbf{V-shape:} Minimum $\eta/s = 0.08$ at $T_c = 0.15$~GeV, rising both below and above $T_c$
    \item[ \textbf{(4)}] \textbf{Low-$T$ slope:}  $\eta/s = 0.08$ at $T_c$, rising below $T_c$, flat above
    \item[ \textbf{(5)}] \textbf{High-$T$ slope:} $\eta/s = 0.08$ at $T_c$, flat below $T_c$, rising above.
\end{enumerate}

\subsection{Scenario II: the \textit{ab initio} nuclear structure in O+O collisions}
\label{sec:methods_nuclear}

Scenario~II applies the same automated pipeline to {O$+$O} collisions at $\sqrt{s_{\mathrm{NN}}}=5.36$~TeV in the 0--5\% centrality. Because the hydrodynamic setup is kept fixed across the four calculations, differences in the final observables reflect the different nuclear-structure
inputs within this controlled model comparison.

The {T\raisebox{-.5ex}{R}ENTo} model~\cite{Moreland2015TRENTo} reads precomputed nucleon positions from HDF5 files generated by the respective nuclear structure calculations and produces transverse entropy density profiles on an event-by-event basis. The workflow automatically parses the multiplicity value embedded in each event header and retains only those events falling within the desired centrality window. For the O+O system at 5.36~TeV, this centrality class corresponds to a total entropy range that yields approximately $dN_{\mathrm{ch}}/d\eta\approx 130$ after hydrodynamic evolution. Each selected {T\raisebox{-.5ex}{R}ENTo} event is converted into a three-dimensional energy density profile by scaling the entropy density with a tuned $K$-factor. The longitudinal profile is constructed using a flat plateau of width $\eta_{\mathrm{flat}}=2.0$ with Gaussian tails of width $\eta_{\mathrm{gw}}=1.8$.



Four distinct \textit{ab initio} descriptions of the $^{16}$O ground state were used as initial-state inputs.
These initial nuclear structures are widely employed in Monte Carlo simulations of O+O collisions. The first is based on nuclear lattice effective field theory (NLEFT)~\cite{k8rb-jgvq}, in which ground-state nucleon configurations are sampled by Euclidean-time projection with auxiliary-field Monte Carlo.
The second and third models are obtained from the projectile generator coordinate method (PGCM) \cite{k8rb-jgvq}: one using a clustered $\alpha$-particle geometry (PGCM clustered) and another using a uniform nucleon distribution (PGCM uniform). The fourth model comes from variational Monte Carlo (VMC) \cite{PhysRevC.96.024326} calculations.

Anisotropic flow coefficients are computed with the two-particle cumulant method for charged hadrons within $|\eta|<0.8$ and
$0.2<p_T<3.0$~GeV/$c$. For each event, we construct
\begin{equation}
Q_n=\sum_{j=1}^{M} e^{in\phi_j},
\qquad
\langle 2\rangle_n =
\frac{|Q_n|^2-M}{M(M-1)},
\end{equation}
where $M$ is the charged-particle multiplicity in the analysis acceptance. The subtraction of $M$ removes self-correlations. The integrated flow is then obtained from the event-averaged two-particle correlator,
\begin{equation}
v_n\{2\} = \sqrt{\langle \langle 2\rangle_n\rangle_{\rm ev}} .
\end{equation}

Statistical uncertainties follow the analysis convention adopted throughout:
integrated $v_n\{2\}$ uncertainties are estimated from five-fold subsampling,
the Pearson correlation coefficients $\rho$ and $\text{corr}(\beta_2, \varepsilon_2)$ from jackknife
resampling, and event-ensemble averages such as $dN/d\eta$ and
$\langle p_T\rangle$ from the standard error of the mean,
$s_X/\sqrt{N_{\rm ev}}$, where $s_X$ is the event-by-event standard deviation.

In addition to the final-state observables, the agent-based analysis evaluates a set of initial-state quantities on the same T\raisebox{-.5ex}{R}ENTo events, including the eccentricities $\varepsilon_2$ and the four-particle cumulant ratio $\varepsilon_2\{4\}/\varepsilon_2\{2\}$. To interpret the O+O comparison, the agent further introduced two geometric diagnostics—the transverse compactness $d_{\mathrm{area}}$ and the event-by-event quadrupole amplitude $\beta_2$—which are defined where they are first used in Eqs.~(\ref{eq:beta2}, \ref{eq:darea}) in Sec.~\ref{sec:results}.



\begin{figure*}[htbp]
\makebox[\linewidth][c]{\includegraphics[width=1.0\linewidth]{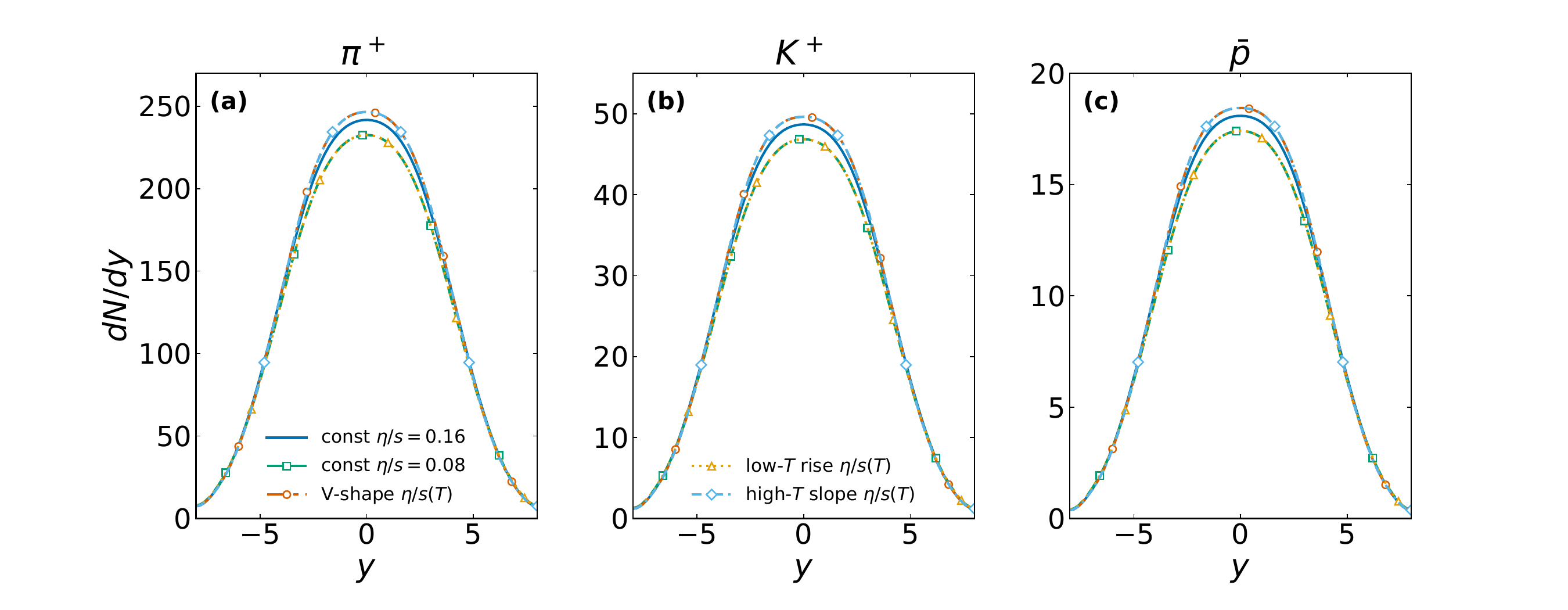}}
\caption{Rapidity distributions $dN/dy$ for (a) $\pi^+$, (b) $K^+$, and (c) $\bar{p}$ under the five $\eta/s$ parametrizations in Scenario~I. 
}
\label{fig:dNdY}
\end{figure*}

\begin{figure*}[htbp]
\centering
\includegraphics[width=0.85\linewidth]{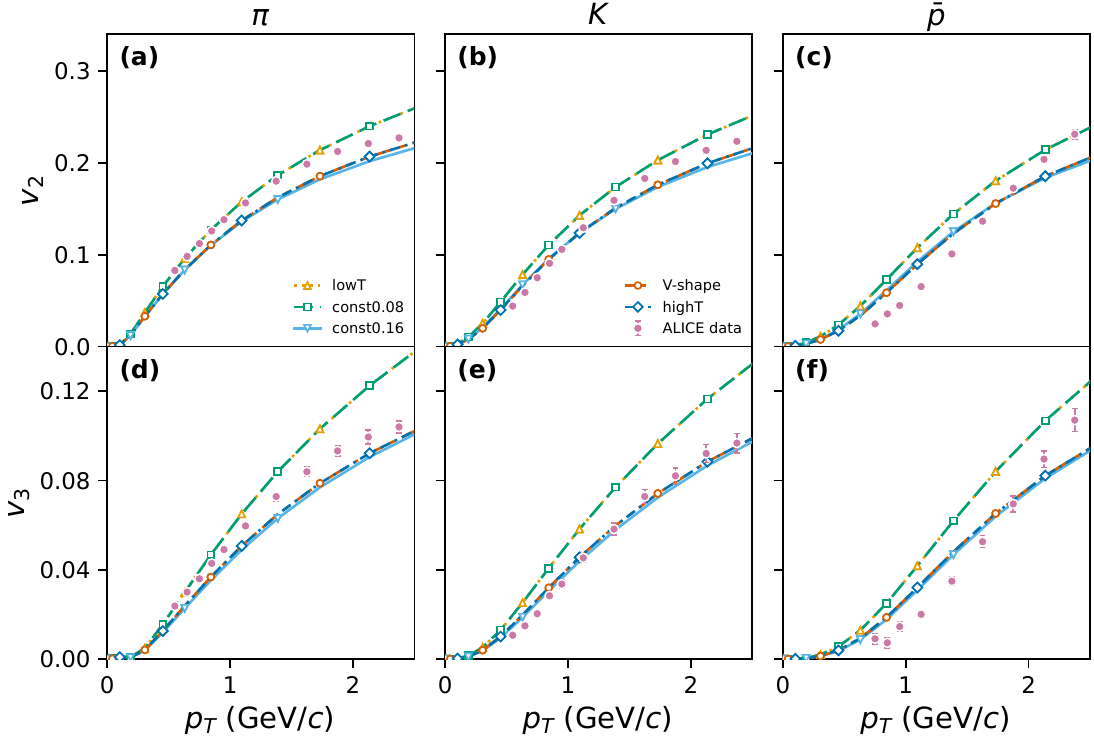}
\caption{Flow harmonics $v_2(p_T)$ (top row) and $v_3(p_T)$ (bottom row) for
$\pi^+$, $K^+$, and $\bar{p}$ under different $\eta/s$ parametrizations in Scenario~I, compared with ALICE measurements for Pb--Pb collisions at $\sqrt{s_{NN}}=5.02$ TeV in the 30--40\% centrality class~\cite{acharya2018anisotropic}. The constant
low-viscosity case ($\eta/s=0.08$) and the low-temperature-slope configuration
produce nearly overlapping, larger flow harmonics, whereas the constant
high-viscosity, V-shape, and high-temperature-slope configurations cluster
together at lower values. }
\label{fig:v2_shear}
\end{figure*}

\section{Results}
\label{sec:results}

\subsection{Scenario I: Effects of temperature-dependent shear viscosity over entropy density}
\label{sec:results_shear}

\subsubsection{Event-averaged observables.}

The agentic pipeline began its analysis of Scenario I by organizing the
event-averaged final-state observables into a coherent physical narrative.
Informed by its knowledge of standard heavy-ion observables, the agent first extracted rapidity distributions $dN/dy$ for $\pi^+$, $K^+$, and $\bar{p}$ across all five viscosity parametrizations, with the results displayed in Fig.~\ref{fig:dNdY}. All five distributions exhibit very similar shapes, peaked symmetrically at midrapidity ($y = 0$), indicating that the longitudinal dynamics of the expanding fireball—governed
by the initial Milne-frame velocity field and the longitudinal
profile—are largely decoupled from the transverse viscous dynamics encoded in
$\eta/s$. The agent therefore focused its subsequent comparison on the
overall normalization and on observables in the transverse sector, where the
sensitivity to viscosity is most pronounced.

Fig.~\ref{fig:dNdY} shows the rapidity distributions of identified particles after
resonance decays for the five $\eta/s$ parametrizations in Scenario~I.
The constant low-viscosity case ($\eta/s=0.08$) and the low-temperature-slope
parametrization give the largest midrapidity yields for all three species,
while the constant $\eta/s=0.16$ case lies between these results and the
V-shape/high-temperature-slope parametrizations, which give the smallest yields. This ordering indicates that the final identified-particle multiplicity is not controlled only by the nominal minimum value of
$\eta/s$, but also by how the temperature dependence modifies the hydrodynamic evolution and the associated entropy production.

Among the three temperature-dependent parametrizations, the rapidity
distributions exhibit a clear separation between the low-temperature-slope case and the other two profiles. The low-temperature-slope parametrization
closely follows the constant $\eta/s=0.08$ result, whereas the V-shape and high-temperature-slope parametrizations form a lower-yield group. Since the
latter two cases share an enhanced high-temperature branch of $\eta/s$, this ordering indicates that the yield normalization is mainly shaped by the
early high-temperature stage of the hydrodynamic evolution. In contrast,
raising $\eta/s$ only in the low-temperature region has little impact on the
final identified-particle multiplicities.


\begin{table}[htbp]
\centering
\caption{Mean transverse momentum $\langle p_T\rangle$ (GeV/$c$) after
resonance decays for the five $\eta/s$ parametrizations (Scenario~I).
Roman numerals denote: I~$=$~const.~$\eta/s=0.16$; II~$=$~const.~$\eta/
s=0.08$;
III~$=$~V-shape; IV~$=$~low-$T$ slope; V~$=$~high-$T$ slope.}
\label{tab:study1b}
\begin{tabular}{@{}lccccc@{}}
\toprule
& I & II & III & IV & V \\
\midrule
$\pi^+$ &$\;\;$ 0.649 $\;\;$& $\;\;$0.642 $\;\;$&$\;\;$ 0.660 $\;\;$ &$\;\;$ 0.642 $\;\;$&$\;\;$ 0.660 \\
$K^+$ & 0.958 & 0.944 & 0.976 & 0.945 & 0.976 \\
$\bar{p}$ & 1.343 & 1.324 & 1.373 & 1.324 & 1.373 \\
\bottomrule
\end{tabular}
\end{table}

\begin{figure*}[htbp]
\centering
\includegraphics[width=\linewidth]{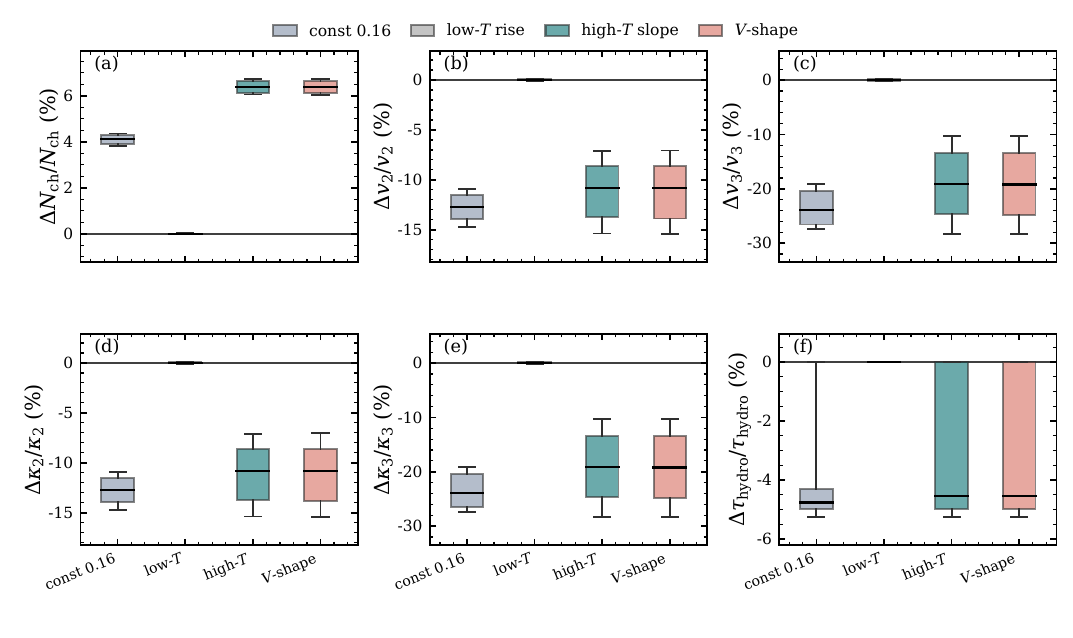}
\caption{Paired-event relative responses $\Delta X/X$ with respect to the $\eta/s=0.08$ constant baseline. Top row (left to right): charged-particle yield $N_{\rm ch}$, elliptic flow $v_2$, triangular flow $v_3$. Bottom row (left to right): hydro response efficiency $\kappa_2=v_2/\varepsilon_2$, hydro response efficiency $\kappa_3=v_3/\varepsilon_3$, hydrodynamic duration $\tau_{\rm dur}$. Boxes span the interquartile range, whiskers indicate the 16th--84th percentile range, and black horizontal lines denote the medians.}
\label{fig:paired_delta}
\end{figure*}

Table~\ref{tab:study1b} shows that the five $\eta/s$ parametrizations
arrange $\langle p_T \rangle$ into three distinct tiers: the V-shaped and
high-temperature-slope profiles yield the highest values, sitting above both
constant-viscosity baselines; the constant $\eta/s = 0.16$ baseline occupies
an intermediate position; and the constant $\eta/s = 0.08$ and
low-temperature-slope profiles cluster together at the lowest values, remaining
nearly indistinguishable across all particle species. The agent further noted
that the enhancement exhibits a mild mass ordering, rising from about $1.7\%$ for $\pi^+$, $1.9\%$ for $K^+$, and $2.2\%$ for $\bar{p}$ relative to the constant
$\eta/s = 0.16$ baseline, and attributed this pattern to stronger radial flow driven by the elevated viscosity at early, high-temperature times.

Elliptic flow $v_2$ most directly encodes the anisotropic geometry of the
initial state, and, recognizing $v_2$ as the most direct probe of initial-state anisotropy, the agent performed a dedicated comparison of its sensitivity to the temperature dependence of $\eta/s$.
Fig.~\ref{fig:eta_s_profiles} displays the five $\eta/s$ parametrizations
used in this scenario, spanning behaviors from flat constant profiles to
sharply V-shaped and monotonically rising forms. Fig.~\ref{fig:v2_shear}
presents $v_2(p_T)$ and $v_3(p_T)$ for $\pi^+$, $K^+$, and $\bar{p}$ together with the
corresponding ALICE measurements~\cite{acharya2018anisotropic}. The constant-viscosity comparison reproduces
the expected trend: reducing $\eta/s$ from $0.16$ to $0.08$ increases $v_2$ near $p_T\simeq 1.1$~GeV/$c$ by about $15$--$16\%$ for all three species. Among the temperature-dependent parametrizations, the same grouping observed in the yield and mean-$p_T$ observables persists. The low-temperature-slope
profile remains nearly indistinguishable from the constant $\eta/s=0.08$
case, whereas the V-shaped and high-temperature-slope profiles stay close to
the constant $\eta/s=0.16$ baseline, despite their different behavior below
$T_c$. This indicates that the high-temperature branch of $\eta/s$ controls
most of the viscous suppression of anisotropic flow, while modifying
$\eta/s$ only in the low-temperature region has little visible impact on
the final differential flow.

Although the parameters are not fine-tuned to experimental data, a rough comparison with ALICE measurements indicates that the measured $p_T$-differential $v_2$ and $v_3$ lie between these two branch curves. In most cases—for instance, $v_2(p_T)$ for kaons and protons, and $v_3(p_T)$ for pions, kaons, and protons—the curves corresponding to constant $\eta/s=0.16$, the V-shaped parametrization, and the high-temperature-slope scenario generally sit closer to the experimental data points. However, the $v_2(p_T)$ for pions is notably closer to the $\eta/s=0.08$ curve. This discrepancy is mainly attributed to the initial conditions and the absence of a hadronic cascade module in the present workflow, which particularly affects the $v_3$ to $v_2$ ratio and the proton $v_n$. Since this paper serves merely as a proof-of-principle test of the LLM-agent workflow, we do not attempt to quantitatively reproduce the experimental data, as that would depend on intricate parameter combinations across a large parameter space. We defer this task to a future Bayesian global analysis employing an agentic workflow for high-energy nuclear physics.


\subsubsection{Event-by-event viscosity response.}

To isolate the viscosity-induced modification of the hydrodynamic response
from initial-state geometry effects, the agent independently devised a paired-event analysis: the same ensemble of 200 T\raisebox{-.5ex}{R}ENTo initial conditions was
evolved under all five $\eta/s$ profiles, so that differences in the final-state observables reflect changes in the viscosity profile alone. The constant low-viscosity case, $\eta/s=0.08$, is used as the reference because it gives the weakest viscous damping among the constant-viscosity baselines. The relative response is defined for each event as
\begin{equation}
\frac{\Delta X}{X} = \frac{X({\rm profile}) - X({\rm const}\,0.08)}{X({\rm const}\,0.08)},
\end{equation}
which quantifies how a given observable changes when only the viscosity profile is modified.

Fig.~\ref{fig:paired_delta} summarizes the paired-event relative responses for
six quantities: the charged-particle yield $N_{\rm ch}$, the integrated
elliptic and triangular flows $v_2$ and $v_3$, the hydrodynamic response
efficiencies $\kappa_2 = v_2/\varepsilon_2$ and
$\kappa_3 = v_3/\varepsilon_3$, and the hydrodynamic duration
$\tau_{\rm dur}$. The agent's paired-event comparison revealed a consistent
and physically interpretable hierarchy. For the constant $\eta/s = 0.16$
profile, the median shifts relative to the constant $\eta/s=0.08$ reference
are $\Delta N_{\rm ch}/N_{\rm ch}=+4.1\%$,
$\Delta v_2/v_2=-12.9\%$, and $\Delta v_3/v_3=-24.4\%$. The V-shaped and
high-$T$ slope profiles show nearly identical responses:
$\Delta N_{\rm ch}/N_{\rm ch}\approx +6.4\%$,
$\Delta v_2/v_2\approx -11.4\%$, and
$\Delta v_3/v_3\approx -19.1\%$, confirming the near-degeneracy already
observed in the event-averaged comparison. The low-$T$ slope profile produces
only marginal deviations from the reference in all three quantities,
reinforcing the conclusion that the sub-$T_c$ behavior of $\eta/s$ has
negligible leverage on the integrated observables considered here in the present setup.

A second finding from the agent's paired-event comparison is that $v_3$ is substantially more fragile than $v_2$ under viscous modification: the relative
suppression of $v_3$ is about $1.7$--$1.9$ times larger than that of $v_2$
across the non-trivial profiles. This hierarchy is consistent with the general
expectation that higher-order flow harmonics, which encode finer spatial
structure, are more efficiently damped by dissipation, motivating the use of
$v_3$ as a differential probe of the viscosity profile in future experimental
analyses. The response efficiencies $\kappa_2$ and $\kappa_3$, which measure
the intrinsic conversion of spatial eccentricity into momentum anisotropy and
are insensitive by construction to changes in initial geometry, exhibit the
same suppression pattern as $v_2$ and $v_3$, directly reflecting altered
hydrodynamic damping rather than changes in the initial state. The hydrodynamic duration included in Fig.~5 provides a first dynamical diagnostic for this response hierarchy. Profiles with enhanced
high-temperature viscosity shorten the hydrodynamic lifetime while suppressing
$\kappa_n=v_n/\varepsilon_n$, consistent with stronger early-time viscous
damping and a reduced efficiency for converting the initial eccentricity into
final momentum anisotropy. This paired-event result motivates a more focused
look at intermediate hydrodynamic diagnostics, which can clarify why the
response efficiency varies from event to event.

\subsubsection{Dynamical evolution and intermediate hydrodynamic features.}

To understand why the response efficiencies $\kappa_n=v_n/\varepsilon_n$ vary from event to event,the agent extended the
analysis from final-state observables to intermediate hydrodynamic diagnostics
stored in the CLVisc \texttt{bulkinfo.h5} output. We focused on two compact
quantities: the hydrodynamic duration $\tau_{\rm dur}$, defined as the time
during which the system remains above the freeze-out threshold, and the
time-integrated hot transverse area,
\begin{equation}
A_T^{\rm int}(T>0.2~{\rm GeV})
=
\int d\tau\, A[T(x,y,\tau)>0.2~{\rm GeV}],
\end{equation}
which measures the accumulated transverse space-time volume of matter hotter
than $0.2$~GeV.

The event-level analysis showed that $\kappa_2$ and $\kappa_3$ are not fixed
conversion factors from initial eccentricity to final flow. Instead, they
increase systematically with $\tau_{\rm dur}$ and $A_T^{\rm int}$, indicating
that the initial eccentricity provides only the geometric seed, while the
for $\kappa_3$ than for $\kappa_2$, consistent with the stronger viscous
sensitivity of higher harmonics. These intermediate diagnostics therefore
provide the dynamical link between the paired-event response shown in
Fig.~5 and the agent's interpretation of viscosity-driven flow suppression.
They are used here as exploratory diagnostics rather than as independent
constraints on $\eta/s$.

\subsection{Scenario II: The \textit{ab initio} nuclear structure in O$+$O collisions}
\label{sec:results_nuclear}

\begin{figure*}[htbp]
    \centering
    \includegraphics[width=\linewidth]{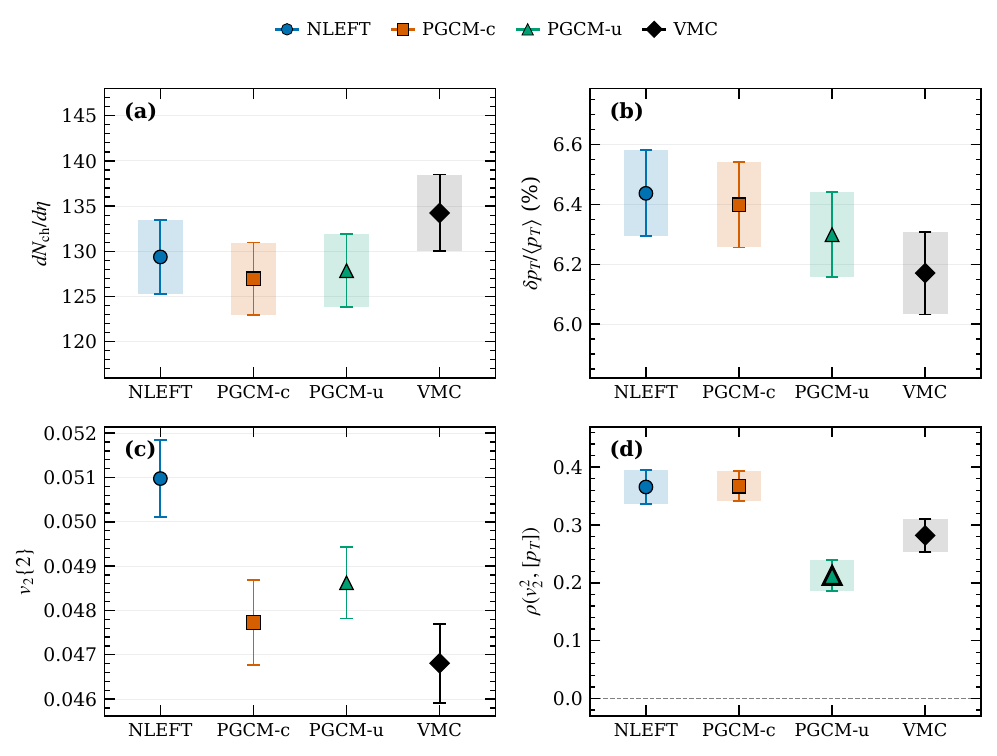}
    \caption{Final-state observables in 0--5\% O+O collisions at $\sqrt{s_{\mathrm{NN}}}=5.36$~TeV for the four $^{16}$O structure models. (a)~charged-particle multiplicity $dN_{\mathrm{ch}}/d\eta$; (b)~relative mean-$p_T$ fluctuation $\delta p_T/\langle p_T\rangle$; (c)~elliptic flow $v_2\{2\}$ ($0.2<p_T<3.0$~GeV/$c$); (d)~elliptic-flow--mean-$p_T$ correlation $\rho(v_2^2,[p_T])$, with PGCM-u the clear outlier. Shaded bands show $\pm1\sigma$ statistical uncertainty.}
    \label{fig:summary}
\end{figure*}

Turning to Scenario II, the agentic pipeline generated 1000 hydrodynamic events for each of the four nuclear structure models in the 0--5\% centrality class of O+O collisions at $\sqrt{s_{\mathrm{NN}}}=5.36$~TeV. Because the hydrodynamic parameters, centrality definition, and analysis cuts were held fixed across all four calculations by the automated workflow design, any systematic differences in the final-state observables can be traced primarily to the nuclear-structure input.   

Fig.~\ref{fig:summary} presents (a) the charged-particle multiplicity $dN_{\mathrm{ch}}/d\eta$, (b) the relative mean-transverse-momentum fluctuation $\delta p_T/\langle p_T\rangle$, (c) the integrated elliptic flow $v_2\{2\}$, and (d) the elliptic-flow--mean-$p_T$ correlation $\rho(v_2^2,[p_T])$ for the four \textit{ab initio} $^{16}$O structure models. Taken together, these four panels suggest that it may be possible to discriminate among the $^{16}$O structure models using these final-state observables. The charged-particle multiplicities are $dN_{\mathrm{ch}}/d\eta \simeq 130$ for all models, consistent with experimental data \cite{Modak:2026ymk,CMS:2025oo_dndeta}.  
NLEFT exhibits the largest values for (b) the relative mean $p_T$ fluctuation, (c) $v_2\{2\}$, and tied with PGCM-c at the top of (d) the $v_2^2$--mean-$p_T$ correlation. VMC yields the smallest mean $p_T$ fluctuation and $v_2\{2\}$. PGCM-c shows slightly larger mean $p_T$ fluctuation and $v_2\{2\}$ than VMC, but follows the same hierarchy, making it difficult to distinguish VMC from PGCM-c (with clustering). In contrast, PGCM-u (without clustering) displays the weakest $v_2^2$--mean-$p_T$ correlation, even though both its mean $p_T$ fluctuation and $v_2\{2\}$ are larger than those of VMC. As a conclusion, at a fixed charged-particle multiplicity, it might be possible to separate NLEFT, VMC, and PGCM-u by their distinct feature combinations. VMC and PGCM-clustered remain degenerate: they are statistically consistent in (b) and (c) and share the same positive initial-state size--shape coupling ( also see Table~\ref{tab:s2initial}), so no single observable cleanly isolates one from the other.

The model ordering of $\rho(v_2^2,[p_T])$ does not follow that of $v_2\{2\}$: rather than the magnitude of the elliptic response, this observable measures how the shape-driven and size-driven responses co-fluctuate event by event, since $[p_T]$ tracks the transverse size of the fireball through radial flow (at fixed multiplicity, a more compact initial state drives stronger radial flow and hence a larger $[p_T]$). The decisive contrast is within the PGCM family: PGCM-clustered and PGCM-uniform share the same many-body framework and differ essentially only in whether $\alpha$-cluster correlations are retained, yet PGCM-uniform gives $\rho(v_2^2,[p_T])=0.21$, about $42\%$ below PGCM-clustered and NLEFT. Removing clustering decouples elliptic shape from transverse size and suppresses the correlation, singling out PGCM-uniform. VMC, by contrast, gives an intermediate $\rho(v_2^2,[p_T])=0.28$ but retains the same \emph{positive} initial-state size--shape coupling as the clustered models (also see Table~\ref{tab:s2initial}); its lower value reflects its smaller overall elliptic flow, not a loss of coupling. In $\rho(v_2^2,[p_T])$ the qualitative divide is therefore between the clusterless PGCM-uniform and all clustered/realistic configurations, with VMC and PGCM-clustered staying on the same (coupled) side. Combined with the $v_2\{2\}$ ordering, this leaves three separable groups---NLEFT, PGCM-uniform, and the degenerate VMC/PGCM-clustered pair.

Table~\ref{tab:s2initial} presents the initial-state diagnostics for the four
$^{16}$O models, extracted from the T\raisebox{-.5ex}{R}ENTo initial conditions and from the
sampled nucleon configurations. Taken together, they show that the two
final-state discriminators of Fig.~\ref{fig:summary} arise from two distinct and
separable pieces of nuclear-structure information: the \emph{amplitude} of the
elliptic shape fluctuations, which sets the $v_2\{2\}$ ordering, and the
\emph{size--shape correlation} of the initial fireball, which controls
$\rho(v_2^2,[p_T])$. The amplitude is diagnosed by $\langle\varepsilon_2\rangle$,
$\varepsilon_2\{4\}/\varepsilon_2\{2\}$, and---at the deeper level of the nucleon
configurations---the intrinsic quadrupole amplitude $\beta_2$ (with its relative
width $\sigma(\beta_2)/\langle\beta_2\rangle$ and its event-wise correlation with
$\varepsilon_2$); the size--shape correlation is quantified by
$\rho(\varepsilon_2^2,d_{\mathrm{area}})$. We take the two in turn, and then ask
what places VMC at the bottom of the amplitude ordering.

One open question is why the eccentricity of VMC is the smallest. A previous study by Ref.~\cite{ZHANG2025139322} attributed this to short-range repulsive nucleon--nucleon correlations, which lead to a more uniform distribution of nucleons within confined regions and reduce configuration fluctuations~\cite{Wang_2026}. In contrast, NLEFT and PGCM lack such short-range correlations. During the analysis stage of the workflow, the agent proposed a new observable to explain the small eccentricity of VMC: the intrinsic quadrupole amplitude $\beta_2$ for each sampled nucleon configuration, defined in the standard Bohr–Mottelson convention as
\begin{equation}
\beta_2 = \sqrt{\frac{5\pi}{9}}\frac{\lambda_{\max}}{A R_0^2},
\qquad
Q_{ij}=\sum_{k=1}^{A}\left(3r_{ki}r_{kj}-r_k^2\delta_{ij}\right),
\label{eq:beta2}
\end{equation}
where $Q_{ij}$ is the traceless quadrupole tensor of the nucleon coordinates,
$\lambda_{\max}$ its largest eigenvalue, $R_0=1.2A^{1/3}$~fm, and $A=16$.
Since $^{16}$O is a few-body system, $\beta_2$ here measures an event-by-event shape fluctuation that resembles deformation.
For ground-state deformation, $\beta_2=0$ corresponds to a spherical nucleus, while a positive $\beta_2$ indicates an ellipsoidal shape.
As shown in Table~\ref{tab:s2initial}, $\beta_2$ exhibits a strong correlation with $\varepsilon_2$, with correlation coefficients close to 0.60 for all four nuclear structures.
Thus, $\beta_2$ closely tracks the ordering of eccentricity: VMC has the smallest mean $\beta_2$ and fluctuation, with $\langle\beta_2\rangle=0.44$ and $\sigma(\beta_2)/\langle\beta_2\rangle=0.39$, compared to $\langle\beta_2\rangle=0.57$--$0.61$ and $\sigma(\beta_2)/\langle\beta_2\rangle=0.45$--$0.50$ for the other three models.
The definition of $\beta_2$ was unfamiliar to us initially. We first thought the agent had invented a new observable (which is different from the explanation of Zhang \textit{et al.}~\cite{ZHANG2025139322}), but later found that this observable was originally introduced in~\cite{PhysRevC.105.014905} to study the relationship between intrinsic quadrupole deformation and eccentricity in central collisions. This demonstrates that the agent possesses broad vision, capable of understanding physical observables defined in other studies and generalizing them to new research contexts such as the present work. 

The agent has defined a new initial state "observable" to explain why PGCM-uniform, despite having an elliptic flow comparable to others,
gives such a suppressed final-state $\rho(v_2^2,[p_T])$. This traces to the
near-absence of an initial-state correlation between geometric eccentricity and transverse
compactness. To quantify the latter, the agent introduced the
compactness diagnostic $d_{\mathrm{area}}$ and correlated it event by event
with $\varepsilon_2^2$,
\begin{equation}
d_{\mathrm{area}}=\sqrt{\frac{N_{\mathrm{part}}}{A_\perp}},\qquad
A_\perp=\sqrt{\langle\tilde x^2\rangle_s\langle\tilde y^2\rangle_s-\langle\tilde x\tilde y\rangle_s^2},
\label{eq:darea}
\end{equation}
where $A_\perp$ is the entropy-weighted transverse area,
$\tilde x=x-\langle x\rangle_s$ and $\tilde y=y-\langle y\rangle_s$, and a
larger $d_{\mathrm{area}}$ (at fixed participant number) marks a more compact
profile. For NLEFT, PGCM-clustered, and VMC the correlation
$\rho(\varepsilon_2^2,d_{\mathrm{area}})$ is positive, of order $+0.08$ to
$+0.12$, whereas for PGCM-uniform it is consistent with zero,
$\rho(\varepsilon_2^2,d_{\mathrm{area}})=-0.006$: in the uniform configuration,
ellipticity and transverse compactness are effectively decoupled. Because the
hydrodynamic evolution carries $\varepsilon_2$ into $v_2$ and transverse
compactness into $[p_T]$, this initial-state decoupling propagates to the final
state and accounts for the suppressed $\rho(v_2^2,[p_T])$. The small
correlation is therefore not a consequence of a small ellipticity, but of a
missing size--shape coupling---tellingly, VMC has the smallest ellipticity of
all yet keeps a sizable positive coupling, which is precisely why it stays with
PGCM-clustered rather than with PGCM-uniform.

\begin{table}[tbp]
\centering
\caption{Initial-state observables for the four $^{16}$O models in 0--5\% O+O collisions. The eccentricity-based quantities are obtained from the {T\raisebox{-.5ex}{R}ENTo} initial conditions, and $\beta_2$ from the sampled nucleon configurations.}
\label{tab:s2initial}
\setlength{\tabcolsep}{4pt}
\begin{tabular}{lcccc}
\toprule
Obs. & NLEFT & PGCM-c & PGCM-u & VMC \\
\midrule
$\langle\varepsilon_2\rangle$ & $0.35$ & $0.33$ & $0.34$ & $0.31$ \\
$\rho(\varepsilon_2^2,d_{\mathrm{area}})$ & $+0.079$ & $+0.117$ & $-0.006$ & $+0.115$ \\
$\varepsilon_2\{4\}/\varepsilon_2\{2\}$ & $0.76$ & $0.74$ & $0.74$ & $0.69$ \\
$\langle\beta_2\rangle$ & $0.60$ & $0.57$ & $0.61$ & $0.44$ \\
$\sigma(\beta_2)/\langle\beta_2\rangle$ & $0.50$ & $0.46$ & $0.45$ & $0.39$ \\
$\mathrm{corr}(\beta_2,\varepsilon_2)$ & $0.62$ & $0.61$ & $0.60$ & $0.60$ \\
\bottomrule
\end{tabular}
\end{table}

\FloatBarrier

\FloatBarrier

\section{Discussion and Conclusions}
\label{sec:discussion}

Extracting QGP transport properties and the imprints of nuclear
structure from heavy-ion data has traditionally relied on
labor-intensive, manually configured parameter scans, which limit the dimensionality of the parameter space that can be explored. To test whether this process can be automated, we encoded operational knowledge of the CLVisc $(3{+}1)$D viscous hydrodynamics code, together with heavy-ion analysis conventions, into a structured SKILL framework, and
let an LLM agent drive the full workflow---designing the scan, running the simulations, computing observables, and interpreting the results. We applied the resulting pipeline to two deliberately distinct problems within the same infrastructure: a scan of five temperature-dependent
shear-viscosity parametrizations $\eta/s$ in {Pb$+$Pb} collisions at$\sqrt{s_{\mathrm{NN}}}=5.02$~TeV (Scenario~I), and a controlled comparison of four \textit{ab initio} $^{16}$O structure models in O+O collisions at $5.36$~TeV (Scenario~II).

In Scenario~I, the agent established that the high-temperature branch of
$\eta/s$ governs the viscous suppression of bulk and flow
observables, while the low-temperature slope below $T_c$ has little leverage; 
In Scenario~II, the four $^{16}$O structure models separate into only three distinguishable groups: VMC and PGCM-clustered remain degenerate and together yield the smallest integrated $v_2$, set by their weak event-by-event quadrupole fluctuations; PGCM-uniform is singled out by an anomalously low elliptic-flow--mean-$p_T$ correlation $\rho(v_2^2,[p_T])$, reflecting a decoupling of ellipticity and transverse compactness in its initial state; and NLEFT lies at the opposite extreme. O+O collisions thus act as a selective probe of $^{16}$O ground-state geometry.

Beyond the specific physics, the study shows that an autonomous agent
can do more than execute a workflow: guided by literature-informed
domain knowledge, it selected the physically relevant observables and
interpreted their model dependence without explicit instruction, turning
the viscosity profile and the nuclear-structure input into quantities
that can be scanned rather than assumed. The O+O results, in particular,
strengthen the case that small collision systems are sensitive probes of $^{16}$O ground-state geometry, with the potential for precision discrimination once the medium dependence is controlled, motivating dedicated O+O running at the LHC.

Several limitations qualify these conclusions. The present simulations include resonance feed-down from post-freeze-out decays, but do not include a hadronic transport afterburner. Effects from hadronic rescattering, resonance regeneration or absorption, and late-stage kinetic evolution are therefore not accounted for.
In Scenario~II the hydrodynamic medium is held fixed, so the observed
model orderings should be regarded as qualitative until they are tested
against variations of the transport coefficients, centrality
definitions, and analysis cuts, and against a fuller assessment of statistical and model uncertainties beyond the present
$\sim\!1000$-event samples. Finally, $\beta_2$ serves as a geometric
diagnostic rather than a first-principles explanation: the microscopic origin of VMC's small flow---whether framed as reduced shape
fluctuations or as short-range repulsive correlations---remains open. On the methodological side, the agent operates within the domain knowledge encoded in the SKILL framework, and its physical interpretations still require expert verification; the present study therefore establishes feasibility and physical soundness rather than fully autonomous, unsupervised discovery.

These limitations map directly onto future work. Incorporating a
hadronic afterburner stage---straightforward within the present
architecture---would yield experimentally comparable spectra and
feed-down--sensitive observables, while varying the medium parameters
and analysis choices would establish the robustness of the
nuclear-structure orderings. More differential probes could sharpen the
discrimination; symmetry-plane correlations, shown to retain
nuclear-geometry sensitivity in O+O and Ne+Ne collisions~\cite{CMS:2025oo_flow},
are one promising example. The standardized, high-dimensional datasets
the pipeline produces are well suited to machine-learning-- or
LLM-assisted searches for such discriminants, and the framework extends
directly to other systems (Ne+Ne, Ar+Sc, Pb+Pb) and centrality classes.

Beyond these refinements within hydrodynamic model scans, the framework
is readily extensible to other hydrodynamics-related problems by
incorporating the corresponding SKILLs. One natural direction is the
spin polarization of hyperons induced by the large initial orbital
angular momentum of noncentral
collisions~\cite{Liang:2004ph,Liang:2004xn,STAR:2017ckg}, whose
quantitative description likewise rests on hydrodynamic simulations
(see, e.g., Refs.~\cite{Fu:2021pok,Becattini:2021iol,Yi:2021ryh,Becattini:2024uha}
and references therein). A second is to embed the present pipeline
within Bayesian analyses of the global QGP properties, using the agent
to orchestrate the large ensembles of runs that such inference requires.
We leave these extensions to future work.

More broadly, this work demonstrates that agentic AI can act not merely
as an automation layer but as an active participant in the scientific
loop, mapping theoretical parameters to observables in a systematic,
reproducible, and scalable way. As such pipelines mature, they offer a
practical route toward AI-assisted discovery across the parameter-rich
landscape of QGP and nuclear-structure physics.

Beyond those specific results, the work makes a broader methodological point. An LLM agent equipped with domain knowledge can autonomously select physically meaningful observables, interpret them in their proper theoretical context, and formulate new quantities to explain unexpected patterns---as when it flagged the event-by-event variation of the response efficiency $\kappa_n=v_n/\varepsilon_n$ as statistically robust and traced it to intermediate hydrodynamic diagnostics, a line of reasoning impractical to pursue manually across hundreds of events. Such automated, GPU-accelerated model scans are becoming a prerequisite at the interface of nuclear structure and heavy-ion physics, and the standardized, high-dimensional datasets they produce are naturally suited to machine-learning or LLM-guided searches for novel observables and correlations. As the interplay between low-energy nuclear structure and high-energy collisions deepens, such data-driven frameworks will be essential for turning small collision systems into precision imaging tools for nuclear structure and QGP transport.

\begin{acknowledgments}
This work is supported by NSFC under Grant No. 12535010, No. 12435009, No. 12075098, and No. 12135011 and by the National Key Research and Development Program of China under Contract No. 2022YFA1605500, by the Chinese Academy of Sciences (CAS) under Grants No. YSBR-088. Computations in this work were carried out at the Nuclear Science Computing Center at Central China Normal University (NSC3). The CLVisc SKILL and data are generated by kimi-cli agent. The authors are responsible for the results and physical explanation.
\end{acknowledgments}

\nocite{CMS:2025oo_flow}
\bibliography{references}

\end{document}